\documentclass[12pt,a4paper]{article}

\usepackage{fullpage,amsmath,amssymb,amsfonts,epsfig,setspace,apacite}
\usepackage{subfigure}

\psfigdriver{dvips}

\begin{document}

\title{\begin{bfseries}{Correlation, selection and the evolution of species networks}\end{bfseries}}

\author{Simon Laird\footnotemark[1], Henrik Jeldtoft Jensen\footnotemark[1] \footnotemark[2]}

\maketitle

\begin{abstract}
We use a generalised version of the individual-based Tangled Nature model of evolutionary ecology to study the relationship between ecosystem structure and evolutionary history. Our evolved model ecosystems typically exhibit interaction networks with exponential degree distributions and an inverse dependence between connectance and species richness. We use a simplified network evolution model to demonstrate that the observed degree distributions can occur as a consequence of partial correlations in the inheritance process. Futher to this, in the limit of low connectance and maximal correlation, distributions of power law form, $P(k){\propto}1/k$, can be achieved. We also show that a hyperbolic relationship between connectance and species richness, $C{\sim}1/D$ can arise as a consequence of probabilistic constraints on the evolutionary search process. 
\end{abstract}

\fnsymbol{footnote}
\footnotetext[1]{Department of Mathematics, Imperial College London,  South Kensington campus,  London SW7 2AZ, U.K.}
\footnotetext[2]{Author for correspondence (h.jensen@imperial.ac.uk)}

\section{Introduction} \label{sec.intro}

The network topology of community structure is an important property of an ecosystem. It plays an integral role in system dynamics, influencing future behaviour whilst emerging as a consequence of dynamical history. This duality of cause and effect exemplifies the nonlinearities inherent to such systems and leads to differing theories about the origins of observed network properties. At least since May's work in the seventies \cite{may74:stab} there have been attempts to explain the degree distributions and connectances of ecological networks by reference to stability issues \cite{mcca00:dive}\cite{pimm91:food}\cite{warr94:maki}\cite{mcca00:dive}. Here we focus on the the role played by correlated inheritance in the structuring of the interaction network and elucidate such effects by presenting two related models of evolutionary ecology.

The first, the correlated Tangled Nature model, is an individual-based model of co-evolution in which the reproduction rate of the individual depends on interactions with co-existing types. Mutations move the population around in trait space, but an offspring that differs by only a few mutations from the parent experiences a set of interactions very similar to that of the parent. The correlation in this model allows species to emerge as localised clusters of individuals in trait space. 

The second is a simplified version of the first and consists of a dynamical network with an invariant node number. At each time step a randomly selected node is removed from the network and replaced by a duplicate of another randomly selected node. During the duplication process the edges are inherited with a probability that ultimately determines the parent-offspring similarity. 

The correlated Tangled Nature model produces exponential degree distributions and we attribute this outcome to correlations in the inheritance process. We use the network model to demonstrate that correlated speciation dynamics can lead to such distributions and, additionally, to other forms observed in real ecological systems. Depending on the level of correlation, the network model distributions take binomial, exponential-like and power law forms with the latter tending to $P(k)=1/k$ in certain limits. Interestingly, the network model dynamics emulate ecological processes without any net growth in nodes nor explicit preferential attachment; aspects which are integral to many power-law network dynamics.

The ensembled data from the Tangled Nature simulations also present another curious network phenomenon: an inverse relationship between connectance and the number of species. This form of dependence has been observed in ecological field data; predominantly food web structured but also mutualistic networks \cite{cohe84:trop}\cite{oles02:geog}\cite{mont03:topo}\cite{basc03:nest}. In a generalised interaction network this relation is thought to represent a delineation of stable and unstable regions of network structure \cite{may74:stab}\cite{mcka00:mean}. Here, we formulate an alternative line of reasoning based on a balance of selection forces and the probabilistic limitations of phase space searching. A simple functional form is derived and then compared to data obtained from the simulations of the Tangled Nature model.

In the following section we define the correlated Tangled Nature model and present the results regarding the degree distributions. To provide insight into these results and propose a causal mechanism for the distributional forms, we introduce the network evolution model in section \ref{sec.netevomodel}. Returning to the Tangled Nature model, in section \ref{sec.CVD} we present the second set of results concerning the inverse relationship between connectance and the number of species. A probabilistic model is then used to explain the phenomenon.  
 
\section{Correlated Tangled Nature model}
\label{sec.ctana}
This version of the Tangled Nature model succeeded its predecessor \cite{chri02:tang}\cite{hall02:time}\cite{diCo03:tang}\cite{ande05:netw} by introducing correlations into the inheritance process. Here, we shall present the construct of the model whilst summarising the correlative effects. We refer the reader to the earlier paper for a more detailed explanation of how this is achieved \cite{lair05:tang}.
 
Individuals, $\{\alpha, \beta, ...\}$ are described by phenotype vectors of $L=16$ traits, $\bf{T}^{\alpha}=(\it{T}^{\alpha}_{1},\it{T}^{\alpha}_{2},...,\it{T}^{\alpha}_{L})$, with each trait taking a value from the periodically bounded range, $[0,99999]$. We then define a completely specified interaction matrix ({\bf J}-matrix) that accounts for all interactions between any possible phenotypes. Ultimately only a small number of distinct phenotypes will exist in our system and their interactions will be described by a small subset of the elements of this complete matrix. A proportion, $\theta$, of the entries of the complete {\bf J}-matrix, ${\rm{\bf J}}({\rm{\bf T}}^{\alpha}, {\rm{\bf T}}^{\beta})$, are assigned normally distributed values that have local correlations within the matrix structure. All other values of the remaining proportion of the matrix, $1-\theta$, are assigned zeroes which are treated as a lack of interaction between the two relevant phenotypes. The result is that given a mutation of one trait value we have an exponential decay in the correlation between parent and offspring interaction set values that is dependent upon the distance mutated in the trait value,

\begin{equation}
{\bf{C}} ({\bf{J}}({\bf{T}}^{\alpha},{\bf{T}}^{\gamma}),{\bf{J}}({\bf{T}}^{\beta},{\bf{T}}^{\gamma})) = exp[-{\Delta}({\bf{T}}^{\alpha},{\bf{T}}^{\beta}) /\xi]\,\,{\in(0,1]},
\label{eq.correlation} 
\end{equation}

Here, $\xi$ is the correlation length and ${\bf{C}} ({\bf{J}}({\bf{T}}^{\alpha},{\bf{T}}^{\gamma}),{\bf{J}}({\bf{T}}^{\beta},{\bf{T}}^{\gamma}))$ is the correlation between the interaction strengths of two phenotypes $\alpha$ and $\beta$ when each are interacting with a third, $\gamma$. This measure shall be abbreviated to ${\bf{C}} ({\bf{T}}^{\alpha},{\bf{T}}^{\beta})$ hereafter. If $\beta$ is the mutated offspring of $\alpha$, then it inherits the same phenotype vector but with a small random shift in one of the trait values. The function, $\Delta({\bf{T}}^{\alpha},{\bf{T}}^{\beta})={\mid}\it{T}^{\alpha}_{mut}-\it{T}^{\beta}_{mut}{\mid}$ represents this single trait shift (when two similar phenotypes with multiple trait differences are considered, this function becomes more complicated; we refer the reader to the original paper for details \cite{lair05:tang}). Importantly, interacting uncorrelated phenotypes take values ${\rm{\bf J}}({\rm{\bf T}}^{\alpha}, {\rm{\bf T}}^{\beta})$ and ${\rm{\bf J}}({\rm{\bf T}}^{\beta}, {\rm{\bf T}}^{\alpha})$ that are uncorrelated thus permitting any interaction type (predator-prey, mutualistic etc.) to exist in principle. 

We initiate the system with a set of individuals which are assigned random phenotype vectors. At each subsequent timestep a randomly selected individual is annihilated with probability $P_{kill}=0.2$ whereby it is removed from the system and the single resource unit associated with it is returned to the resource bath, $R(t)$. During the same timestep, another individual is randomly selected to reproduce with probability $P_{repro}$. This value is determined through the use of a weight function,

\begin{equation}
\it{H} ({\bf T}^{\alpha},\it{t}) = 
  a_{1} \frac{\sum_{{\bf T}\in {\cal T}} {\bf J} ({\bf T}^{\alpha}, {\bf T}) n({\bf T},t)} {\sum_{{\bf T}\in {\cal T}} {\bf C} ({\bf T}^{\alpha}, {\bf T}) n({\bf T},t)} 
- a_{2} \sum_{{\bf T}\in {\cal T}} {\bf C} ({\bf T}^{\alpha}, {\bf T}) n({\bf T},t) 
- a_{3} {\frac{N(t)}{R(t)}}.\label{eq.hamiltonian3}
\end{equation}

This is monotonically mapped to the interval [0,1], appropriate for a probability measure, by using the following function,

\begin{equation}P_{repro}=\frac{exp[{\it{H}}({\bf{T}}^{\alpha},{\it{t}})]}{1+exp[{\it{H}}({\bf{T}}^{\alpha},{\it{t}})]}.\label{eq.compfermi}
\end{equation}

The sums of Eq.(\ref{eq.hamiltonian3}) are made over the points in phenotype space, ${\cal T}$, and the occupancies (population associated with each phenotype), $n({\bf T},t)$ are used to account for the multiplicity of individuals with the same phenotype vector. We consider here a well mixed system of constant spatial size, although spatial extent is not explicitly considered. For clarity we reiterate at this point that the phenotype space is a pre-defined, complete set of all possible phenotypes and it is evolution and contingency that select the actualised phenotypes in the evolved system. The {\bf{J}}-matrix is similarly a pre-defined complete set of all possible interactions for all possible phenotypes that may exist {\emph{in potentia}}.

The correlation measure is used in the first term of Eq.(\ref{eq.hamiltonian3}) to restrict the impact of the interaction sum. It represents the fact that interactions are shared amongst members of the same species. For example, a tree may provide a volume of fruit to a solo member of a species but the provision must be shared with reduced efficacy if there are many members. So the overall effect of this denominator, on an individual, is to dampen its interaction sum as a whole thus representing the effect of distributing all interaction effects amongst the individual's own species members. Another example could be a wildebeest in proximity to a lion. The negative predatory effect (or predation probability) of the single lion on that specific wildebeest is decreased if there are many wildebeest about to select from. The interaction is damped by the presence of other members of the individuals own species. This aspect is not ubiquitous in species interactions but does feature in many cases. In recognition of the phenotypic variation inherent in a species, the sum over the correlation values, ${\rm{\bf C}}({\rm{\bf T}}^{\alpha}, {\rm{\bf T}}^{\beta})$, accounts for the fact that species members have different but similar phenotypes. This can be seen as a species description in itself. 

When referring to species in this particular model we specifically mean the wild-type diversity in the sense that the mutant cloud is neglected in this categorisation. But, all interaction effects from all extant phenotypes, mutants included, are accounted for in every interaction sum. The mutant cloud itself is in fact very sparse as we have elected to use a low mutation rate, $P_{mut}=0.0002$. As a consequence, the phenotype distribution is essentially a set of delta points of high occupancy with infrequent mutants existing with low occupancy. This makes the recognition of the wild-type diversity very simple as each species is massively dominated by the population of the wildtype.

The second term of the weight function represents intra-specific competition and uses the same correlation measure as before. Similar but distinct phenotypes are likely to be in competition for resources, space etc. that are specific to their niche. The correlation measure accounts for this similarity. 

The third term represents competition for a conserved vital resource that all phenotypes require for survival. Any successful reproduction event produces an offspring that assumes a unit of resource from the bath, $R(t)$. The conservation requirement, $R(t)+N(t)=constant$, where $N(t)$ is the global population of the system, means we have a carrying capacity for the system as a whole. It's functional form represents the number of system members competing per unit of available resource. 

The parameters, $a_{1}=0.5$, $a_{2}=0.01$, $a_{3}=0.2$ are the selection, conspecific competition and resource competition parameters respectively. These are subjectively chosen to allow interaction controlled dynamics and a sufficient number of species to develop. A value of $\theta=0.05$ is used throughout the simulations.

Upon successful reproduction, the individual produces a single offspring that assumes one unit of resource from the resource bath. As the bath depletes the probability of reproduction for any phenotype is reduced due to the third term in Eq.(\ref{eq.hamiltonian3}). The offspring phenotype is identical to that of the parent (clonal) unless, with probability $P_{mut}=0.0002$, a mutation occurs. This is effected by shifting a randomly selected trait value by a random amount ${\psi}$, that is gaussian distributed with $\mu=0$ and ${\sigma}=\xi$, where $\xi=250$ is the correlation length of the phenotype space. Any mutated offspring will have an interaction set that is similar but not identical to its parent in accordance with the correlation.

\section{Results: degree distribution} \label{sec.model}
\begin{figure}
[ptb]
\begin{center}
\epsfig{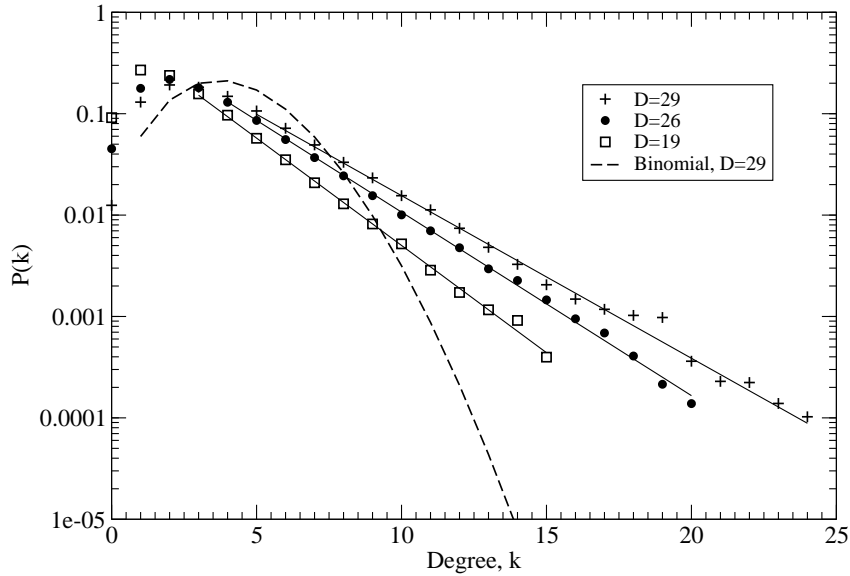}
\caption{\small{Degree distributions for the Tangled Nature model simulations. Shown are ensemble averaged data taken from all networks with number of species, $D=\{19,26,29\}$ over 50 simulation runs of $10^6$ generations each. The exponential forms are highlighted by comparison with a binomial distribution of $D=29$ and equivalent connectance, ${C}{\simeq}{0.145}$ to the simulation data of the same number of species.}}\label{fig.expdeg}
\end{center}
\end{figure}

The degree distribution, $P(k)$ represents the probability of a randomly selected species having $k$ interactions with other members of the system. Given a network where edges are randomly distributed with identical probability, the degree distribution takes binomial form. Many ecological networks present distributions that are far from binomial, assuming forms more appropriately described as exponential or power law \cite{dunn02:food}\cite{jord03:inva}. The version of the Tangled Nature model that incorporates correlated inheritance produces exponential distributions in nearly all cases (Fig.\ref{fig.expdeg}).

 There are theoretical reasons why these forms, particularly the power law, should offer stability to such systems \cite{albe02:stat}, so selection can be proposed as a causal effect for their observed existence. In a nutshell, the network structure permits a greater degree of resilience to random species extinctions. The stability arguments are certainly valid but it may actually be the case that the distributional forms appear as a consequence of the internal dynamics. When we perform simulations with random inheritance the degree distributions revert back to binomial form. This implies that the correlations are an essential requirement for constructing our exponential networks. To support this theory we shall now present a network model that produces a range of non-binomial distributions through correlated dynamics, without any form of selection.
 
\section{Network evolution model} \label{sec.netevomodel}

We consider a system with a fixed number of species, $D$, each defined by generalised interactions with subsets of the other system members. Self-connections are excluded here and as the interaction types are not explicitly considered, the connections are regarded as undirected. This represents a simplified species interaction network that, in principle, embodies interaction types such as mutualism in addition to the usual food-web based predator-prey relationships. As a result, the networks we consider here are not expected to assume non-cyclic tree structures nor the stratified trophic levels associated with resource flow.

We now initiate dynamics representing extinction and correlated speciation with the constraint of an invariant species number. Newly speciated members are seen to supersede extinct members without any implication of cause nor effect. This invariance of the species number can be treated as a consequence of a carrying capacity and whilst simplistic is a reasonable approximation to an ecosystem.
At a timestep, we randomly select with uniform probability a species and all its associated edges to be deleted from the network. A second {\emph{parent}} species is then randomly selected from the remaining $D{-}1$ and is duplicated in the form of a {\emph{daughter}. All species connected to the parent are now given connections to the daughter with probability $P_{e}$. All species unconnected to the parent are given connections to the daughter with probability, $P_{n}$. The determination of an edge between the daughter and parent is made with probability $P_{p}$.

At mean-field level the system admits a single attracting fixed point in the connectance, $\bar{C}{\to}C_{0}$, which is determined by the duplication probabilities (Eq.\ref{eq.theta}). This can be shown when considering the following evolution equation for the number of network edges,
\begin{eqnarray}
E_{t+2} &=& E_{t}-\bar{k}_{t}+{P_{e}}\bar{k}_{t+1}+{P_n}(D-2-\bar{k}_{t+1})+P_{p},\nonumber\\
\bar{k}_{t} &=& \frac{2E_{t}}{D},\,\,\,
\bar{k}_{t+1} = \frac{2}{D-1}(E_{t}-\bar{k}_{t}).\label{eq.mf_evo1}
\end{eqnarray}
The second term accounts for the loss of edges when a node with mean degree, $\bar{k}_{t}$ is deleted in stage one. The third and fourth terms then represent the increase in edges incurred at stage two when one of the remaining $D-1$ nodes is duplicated. Imperfect duplication is made possible at this point by the inclusion of the probabilities, $P_{e}$ and $P_{n}$. The fifth term represents the contribution from the daughter-parent connection.

As we only observe the system after every two steps we may rescale the time variable increment, $(t+t'){\to}(t+\frac{t'}{2})$. Thus Eq.(\ref{eq.mf_evo1}) may be expressed as a recursion relation that has a general solution of the form,
\begin{equation}
E_{t} = A(1+\alpha)^{t}-\frac{\beta}{\alpha}.\\
\label{eq.gsol}
\end{equation}
This admits a single fixed point dependent upon the parameter values which can be expressed in terms of the network connectance,
\begin{equation}
{C}_{0}=\frac{P_{n}(D-2)+P_{p}}{D-1-(P_{e}-P_{n})(D-2)}.
\label{eq.theta}
\end{equation}
This point is an attractor for all physically relevant values; ${D}{>}{2}$; ${0}\leq{P_{e}},P_{n},P_{p}\leq{1}$. By using this relation we set the probabilities $P_{e}, P_{n}$ dependently such that $P_{p}{=}C_{0}$, so allowing $P_{d}{\in}[C_{0},1]$ to be used as a control parameter for the fidelity. As a result the dynamics range from fully random, $P_{e}{=}P_{n}{=}P_{p}{=}C_{0}$, generating binomial networks, to fully correlated, $P_{e}{=}1, P_{d}{=}0, P_{p}{=}C_{0}$, where daughter and parent are indistinguishable.

\subsection{Numerical results}
The ensemble averaged degree distributions for various fidelities are shown in Fig.(\ref{fig.id1}). For the uncorrelated, random duplication process, $P_{d}{=}P_{n}{=}C_{0}$ we see binomially distributed networks as expected. As we increase the fidelity, the distributions exhibit longer tails and conform more to exponential than they do binomial. This progression continues until we have a process of perfect duplication at which point the distributions become power-law-like.


\begin{figure}[]
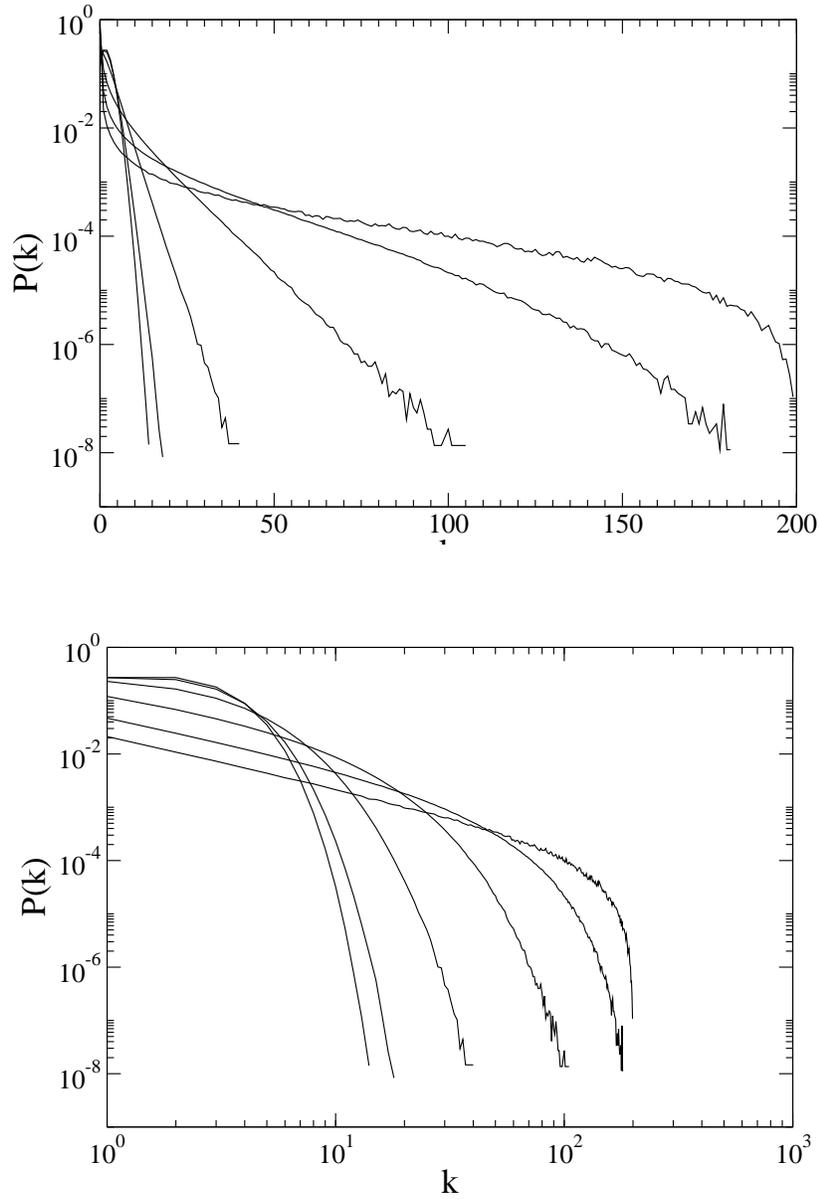

\centering
\subfigure{
\epsfig{file=id_loglin.eps,height=3.0in,angle=0}}
\centering
\vspace{3mm}
\subfigure{
\epsfig{file=id_loglog.eps,height=3.0in,angle=0}}
\caption{\small{Degree distributions for $D=200$, ${C}_{0}=0.01$ systems produced using the imperfect duplication process. From short to long tail we have $P_{d}{=}\{0.01,0.25,0.75,0.95,0.99,0.999\}$.}}
\vspace{3mm} 
\label{fig.id1}
\end{figure}

Shown in Fig.(\ref{fig.pd1}), are the ensemble averaged degree distributions for perfect duplication networks of $C_{0}{=}0.01$ at different system sizes. For small connectance values the distributions conform loosely to a power law. But as we reduce ${C}_{0}$ towards zero, the functional form becomes a discernible power law with exponent, $\gamma{\simeq}1$. The networks achieved here are extremely sparse and the majority of the weight is held in the $P(k{=}0)$ degree. So the form we see here represents (near) scale-free fluctuations. In the limit that ${C}_{0}{\rightarrow}{0}$ a finite sized system asymptotically achieves the absorbing state of zero connectance. As we take the thermodynamic limit in the network size, though, we have a system where $P(k{=}0)$ tends to unity and the remaining distribution achieves power-law form but with $P(k{>}0)$ tending to zero. The simulations suggest that in these limits we have a network of zero connectance with $1/k$ scale-free fluctuations in the degree. It can be verified analytically that this is indeed the case \cite{lair06:stea}.

\begin{figure}[t]
\centering
\epsfig{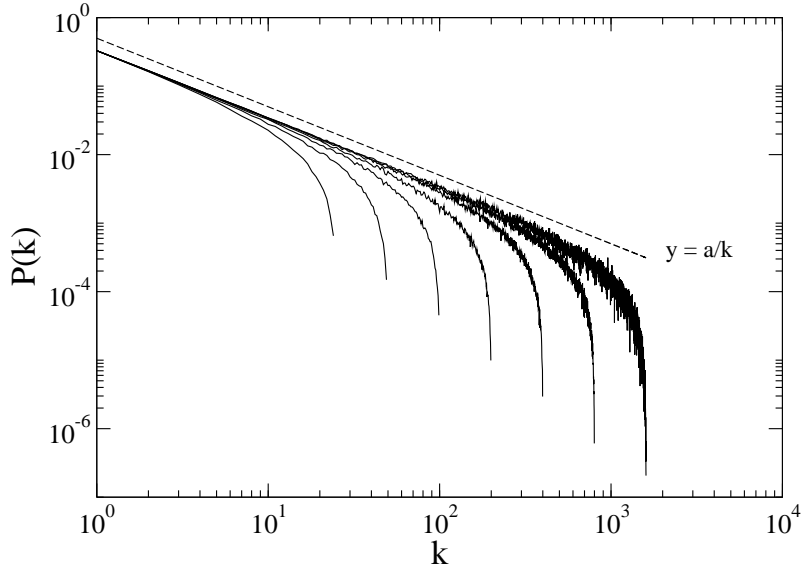}
\caption{\small{Perfect duplication degree distributions for networks with number of species, $D{=}\{25,50,100,200,400,800,1600\}$, ${C}_{0}{=}0.01$. Each has been normalised with the exclusion of the ${k}{=}{0}$ support and rescaled to overlap for visual purposes.}}\label{fig.pd1}
\end{figure}

The power law distribution appears under the conditions of perfect correlation and taking the limit of vanishing connectance (or equivalently, $p_p{\to}0$). Whilst this is not representative of a real system we may relax these conditions whilst still achieving near power law forms. Due to the irresolute nature of field data, observed distributions are naturally indefinite anyway. If we assume though that they do conform to power-laws then it is instructive to compare their exponents. The low value, $\gamma{\simeq}1$ is of significance here as ecological systems with power law distributions generally display such low range exponents,\cite{jord03:inva}\cite{mont03:topo}. Biological systems in general exhibit exponents in the range, $\gamma{\in}(1,2)$, \cite{chun03:dupl} which can be reproduced with duplication-divergence {\emph{growth}} models. Ecological degree distributions find themselves at the lower end of this range, as do those from our model, and the lack of network growth might be an important factor in achieving this.

Field data degree distributions vary in form, with no obvious reason why they should take any one in particular, let alone why multiple forms do actually occur. Our model suggests that these distributions may simply arise as a consequence of correlated dynamics with the nature of the distribution being determined by the level of correlation. The idealised system modelled here fails to account for other defining processes, such as species invasion, and these will have an impact on the correlations. But, ultimately any species derives as an offshoot of an ancestral species so correlations will be inherent to the system even if a level of decorrelation is occurring. 

\section{Connectance-species relationship} \label{sec.CVD}

We now return to the correlated Tangled Nature model and the observed inverse relationship between the connectance, $C$ and number of species, $D$. Connectance is a simple and commonly used measure of a network and it appears explicitly in the May-Wigner criterion of ecosystem stability \cite{may74:stab}. It represents the relative proportion of edges existing in a network when compared to the maximal set of all possible edges. Studies of field data have indicated that the network connectance and the number of species are inversely dependent. The networks created in the correlated version of the Tangled Nature model also exhibited a similar form of dependence (Fig.\ref{fig.cvd_tana}). For the real ecosystem, the {\emph{constant connectance hypothesis} and {\emph{link species scaling law}} \cite{mart92:cons}\cite{cohe90:comm}} have been proposed to describe this relationship but neither viewpoints fully encompass the observed form. The May-Wigner criterion for stability naturally incorporates a $1/D$ dependence for the connectance which is qualitatively similar to the real system. Recently, an analytical population dynamics model that incorporates species invasion has provided an improved parametrised functional fit \cite{mcka00:mean}. The connectance value used in their work was that of the bare network which is analogous to the $\theta$ parameter used in the Tangled Nature model (see section \ref{sec.ctana}). By considering aspects of the Tangled Nature model we propose an alternative mechanism that recognises the role of connectance as a variable in the selection process and so account for the fact that it may deviate from the bare network value.
 
We consider a hypothetical set of all viable organisms each described by distinct phenotypical traits that determine their interactions with other phenotypes. Each organism interacts with a fixed subset of all other possible organisms which as a whole defines a complete species interaction network. An ecosystem represents a sub-network of this complete network where evolution has moulded the phenotype distribution such that a small proportion of the possible organisms are extant. Realistically the strategies employed by organisms can depend upon population frequencies and environmental cues, so defining a complete superset of pairwise interactions is recognised to be an approximation.

We may now ask the question: given a random set of biologically viable organisms what properties would the realised interaction network exhibit? If the complete set $J$ has a connectance $C_{J}$ with a binomially distributed degree then a purely random subset will, on average, have the same connectance and degree distributional form. For the following argument we shall assume that our complete bare network has such properties.

In an evolving system, the effects of selection are partly determined by phenotypic interactions. Positive interactions are beneficial to an organism so we can conceive that selective processes incur an increase in such interactions. Dynamical systems theory suggests that mutualism should not be overly prevalent in community dynamics for stability reasons \cite{may74:stab} but the hypothetical average interaction strength may still tend towards positive values. Mutualism clearly does occur in real systems and interactions described as parasitic may in reality be more commensal than antisymmetric \cite{may74:stab}. The networks exhibited in the Tangled Nature model attained higher than average connectance values with significantly greater mean interaction strengths. This demonstrated that selection processes applied at the level of individuals can cause a global increase in connectance.

We can perceive this increase in edges as a shift upwards from the mean of the binomial distribution. But this shift is counterbalanced by the diminishing binomial probability of achieving higher edge numbers. The phase space of network configurations is heavily dominated by realisations that lie around the binomial mean so a selection-driven search in this space is unlikely to encounter high connectances. We estimate the shift as being a certain proportion, $s$, of the fluctuations one encounters in an unbiased binomial network, $\sigma(D,C_J)$. So in terms of the number of edges, $E$ in the network we estimate,
\begin{eqnarray}
E &=& <E_{J}(D,C_J)> + {s}{\sigma(D,C_J)} \nonumber \\
  &=& E_{m}C_J + {s}[E_{m}C_J(1-C_J)]^{\frac{1}{2}},
\end{eqnarray}
where $E_{m}$ is the total number of edges in a maximally connected network and $<E_J(D,C_J)>$ is the mean number of edges for networks of size, $D$ with edge probability $C_J$. The parameter, $s$ is a measure of selection strength and is assumed to be independent of the system size. The connectance of the evolved network is defined as, $C=\frac{E}{E_{m}}$, which gives,
\begin{eqnarray}
C &=& C_J+{s} \left[ \frac{C_J(1-C_J)}{E_{m}} \right]^{\frac{1}{2}} \nonumber \\
  &=& C_J+{s} \left[ \frac{2C_J(1-C_J)}{D(D-1)} \right]^{\frac{1}{2}}.
\label{eq.cvd}
\end{eqnarray}
Here we have used our definition of the maximal edge set, $E_{m}=\frac{1}{2}D(D-1)$. Other definitions that incorporate self-interaction and directional edges could also be applied.

\begin{figure}[t]
\begin{center}
\epsfig{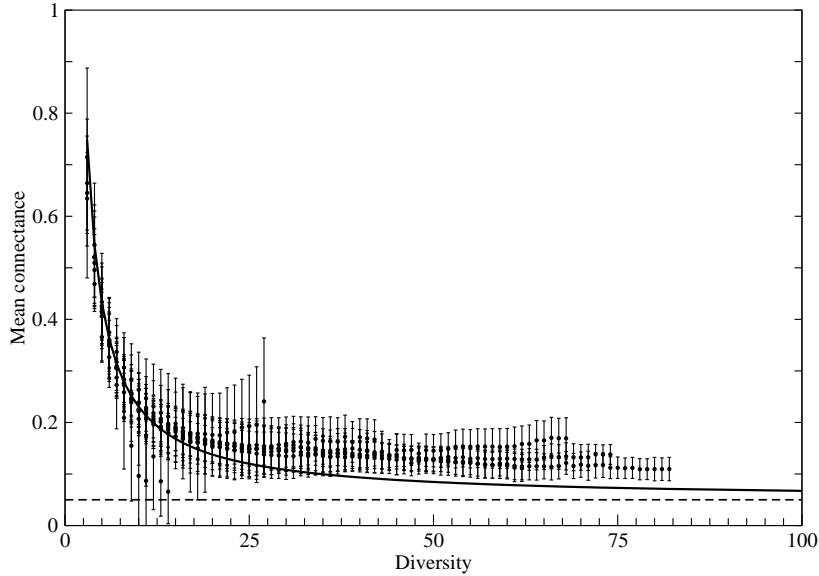}
\caption{\small{Plot of ensemble-averaged mean connectances, $<C>$ against the number of species. Error bars represent the standard error. The lower dotted line marks the null system connectance, ${C}_{J}=0.05$, which the evolved systems clearly surpass. The overlaid functional form is that given by Eq.(\ref{eq.cvd}) using the correct background connectance, ${C}_{J}=0.05$ and with a value of, $s=5.5$ for the selection parameter.}}\label{fig.cvd_tana}
\end{center}
\end{figure}

So we have demonstrated that, with probabilistic restraint, a selective force driving the system to achieve greater numbers of interactions will incur an inverse dependence between connectance and the number of species. The assumption that selection incurs a net increase in interactions is justifiable, but the assumption of a binomial network structure is a significant limitation. The first part of this paper demonstrated how correlations between species can give rise to exponential-like, and power law distributions. The connectance-species relationship might depend on the choice of background distribution so a fuller argument would require consideration of a more appropriate distribution. Even so, the underlying point remains valid. Selection is driving the system towards greater numbers of positive interactions and so higher edge numbers and equivalently higher connectances. But as we increase the number of nodes in our subnetwork the probability of achieving an elevated connectance diminishes, resulting in a decreasing functional form. If we compare Eq.(\ref{eq.cvd}) to the ensemble data acquired from the Tangled Nature model we see that the form is qualitatively appropriate, Fig.(\ref{fig.cvd_tana}). With a background connectance of $C_J=0.05$, the value used in the simulations, the fit is good but not ideal. The simulation networks presented exponential-like degree distributions though so a formulation based upon binomial networks could be responsible for this deviation at higher diversities.

\section{Discussion}
We have shown here that the exponential degree distributions of the correlated Tangled Nature model may be attributed to dynamical rather than selective processes. Our network evolution model dynamics generate distributions ranging from binomial through exponential to power-law which encompasses the Tangled Nature model results and many of the forms observed in real ecological systems. In the case of the power law distribution our exponent ${\gamma}{\simeq}{1}$ compares well with the low values associated with ecological networks that take power law form. The network dynamics are appropriate given the type of system but the model is idealised and ignores other determining factors, such as migration. The random introduction of species acts to decorrelate the system so future work would need to take account of such wider considerations.

Several theories have been proposed to explain the inverse relationship between connectance and the number of species observed in field data ensembles. We have provided reasoning for this phenomenon that has no reliance on stability arguments nor allometric scaling due to resource flow. We argue that evolution pushes an ecology to acquire positive interactions but is restrained due to the improbability of achieving high connectances. A $C{\simeq}{1/D}$ relationship arises as a consequence of this which is in close agreement with the simulation data taken from the correlated Tangled Nature model. The fit is not ideal and field data appears to have a larger exponent than our relation \cite{have92:scal}\cite{pimm91:food}, but our formulation does rely on certain approximations. In particular the assumption of a binomial network is unrealistic. We have demonstrated that dynamics involving correlated inheritance have a strong influence on the form of the degree distributions. An evolutionary search process naturally flows into regions of the network space where the degree distributions are non-binomial and this reflects the inherent correlations. The phase space has non-trivial structure and this can influence the probability of locating higher connectance regions during the search process. Accordingly, the probabilistic reasoning has limitations, but the underlying point still has validity. Future work could attempt to account for the non-binomial networks that feature in ecology and so produce a more appropriate formulation. 

One may ask how our findings can be related to real biological systems. In principle this should be possible within chemostat-based microbial evolution experiments. Two microbial ecosystems with the same initial composition should, as a result of evolution, develop interaction networks with different connectances that are dependent on their species number. The correlations in the inheritance process should ensure they continue to exhibit similar long-tailed functional forms for their degree distributions.



\section*{\small{Acknowledgements}}
We wish to thank Andrew Parry, Kim Christensen, Gunnar Pruessner and the Imperial College Complexity Group for valuable insights. A debt of thanks goes to Andy Thomas for computational support. Simon Laird thanks the EPSRC for PhD funding of this work.


\bibliographystyle{apacite}
\bibliography{paper4}
\end{document}